\documentclass{ws-procs975x65}

\begin{document}

\title{Gravitational Zero Point Energy induces Physical Observables}

\author{REMO GARATTINI}

\address{Universit\`{a} degli Studi di Bergamo, Facolt\`{a} di Ingegneria,\\ Viale
Marconi 5, 24044 Dalmine (Bergamo) ITALY.\\INFN - sezione di Milano, Via Celoria 16, Milan, Italy\\
\email{remo.garattini@unibg.it}}

\begin{abstract}
We consider the contribution of Zero Point Energy on the induced
Cosmological Constant and on the induced Electric/Magnetic charge in absence
of matter fields. The method is applicable to every spherically symmetric
background. Extensions to a generic $f\left( R\right) $ theory are also
allowed. Only the graviton appears to be fundamental to the determination of
Zero Point Energy.
\end{abstract}

\bigskip 

\bigskip \bodymatter

The path integral approach to quantum gravity%
\begin{equation}
Z=\int \mathcal{D}\left[ g_{\mu \nu }\right] \exp iS_{g}\left[ g_{\mu \nu }%
\right]   \label{p01}
\end{equation}%
is a powerful method to study the quantization of the gravitational field,
especially in the context of a WKB approximation on a given background.
Indeed, if one considers a background $\bar{g}_{\mu \nu }$, the
gravitational field splits into $g_{\mu \nu }=\bar{g}_{\mu \nu }+h_{\mu \nu }
$, where $h_{\mu \nu }$ is a quantum fluctuation around the background field
and Eq. $\left( \ref{p01}\right) $ becomes 
\begin{equation}
\int \mathcal{D}g_{\mu \nu }\exp iS_{g}\left[ g_{\mu \nu }\right] \simeq
\exp iS_{g}\left[ \bar{g}_{\mu \nu }\right] \int \mathcal{D}h_{\mu \nu }\exp
iS_{g}^{\left( 2\right) }\left[ h_{\mu \nu }\right] ,  \label{p02}
\end{equation}%
where $S_{g}^{\left( 2\right) }\left[ h_{\mu \nu }\right] $ is the action
approximated to second order. Since the second order action is quadratic in $%
h_{\mu \nu }$, the integration in Eq.$\left( \ref{p02}\right) $ is
straightforward. In absence of matter fields, one may wonder what is the
contribution of quantum fluctuations to the cosmological constant, by
computing the averaged energy-momentum tensor%
\begin{equation}
\left\langle T_{\mu \nu }^{C.C.}\right\rangle =-\frac{2}{\sqrt{-g}}\frac{%
\delta \ln Z}{\delta g^{\mu \nu }},  \label{p03}
\end{equation}%
where $T_{\mu \nu }^{C.C.}$ is the energy momentum tensor related to the
cosmological term. A general solution of Eq.$\left( \ref{p03}\right) $ is a
complicated task, however the Arnowitt-Deser-Misner ($\mathcal{ADM}$)
variables offer a valid example to extract the relevant degrees of freedom%
\cite{ADM}. In terms of these variables, the metric background is%
\begin{equation}
ds^{2}=-N^{2}dt^{2}+g_{ij}\left( N^{i}dt+dx^{i}\right) \left(
N^{j}dt+dx^{j}\right) =-N^{2}dt^{2}+\frac{dr^{2}}{1-\frac{b\left( r\right) }{%
r}}+r^{2}\left( d\theta ^{2}+\sin ^{2}\theta d\phi ^{2}\right) .  \label{ds2}
\end{equation}%
We recognize that the \textit{lapse function} $N$ is invariant and the 
\textit{shift function} $N_{i}$ is absent. To have an effective reduction of
the modes, we consider perturbations of the gravitational field on the
hypersurface $\Sigma \subset \mathcal{M}$. This means that we are
\textquotedblleft \textit{freezing\textquotedblright\ }the perturbation of
the lapse and the shift functions respectively. In summary,%
\begin{equation}
\left\{ 
\begin{array}{c}
g_{ij}\longrightarrow \bar{g}_{ij}+h_{ij} \\ 
N\longrightarrow N \\ 
N_{i}\longrightarrow 0%
\end{array}%
\right. .  \label{mpert}
\end{equation}%
If we indicate with $u_{\mu }$ a time-like unit vector, then it is possible
to prove that\cite{Remo}%
\begin{equation}
\left\langle T_{\mu \nu }^{C.C.}\right\rangle =-\frac{i}{2}g_{\mu \nu }\int 
\frac{d^{4}k}{\left( 2\pi \right) ^{4}}\ln \lambda _{TT}^{2}+\frac{i}{2}%
\left( g_{\mu \nu }+u_{\mu }u_{\nu }\right) \int \frac{d^{3}k}{\left( 2\pi
\right) ^{3}}\ln \lambda _{V^{\bot }}^{2},  \label{LIndA}
\end{equation}%
where $\lambda _{TT}$ are the transverse-traceless (TT)\ eigenvalues of the
following second order differential operator%
\begin{equation}
O^{ikjl}h_{kl}=\left[ \bigtriangleup _{L}^{ikjl}+4R^{ij}g^{kl}+\frac{1}{N^{2}%
}\frac{\partial ^{2}}{\partial t^{2}}g^{ik}g^{jl}\right] h_{kl}=-\left(
\lambda _{TT}^{2}\right) h^{ij}
\end{equation}%
and $\bigtriangleup _{L}$ is the Lichnerowicz operator. From Eq.$\left( \ref%
{LIndA}\right) $, integrating out the time component, the energy density
simply becomes%
\begin{equation}
\frac{\Lambda }{8\pi G}=-\frac{i}{2}\left[ \int \frac{d^{4}k}{\left( 2\pi
\right) ^{4}}\ln \lambda _{TT}^{2}\right] =-\frac{1}{2}\sum_{i=1}^{2}\int 
\frac{d^{3}k}{\left( 2\pi \right) ^{3}}\sqrt{\lambda _{i}^{2}\left(
\left\vert \vec{k}\right\vert \right) }.  \label{Lind}
\end{equation}%
The same expression can be obtained by computing%
\begin{equation}
\frac{1}{V}\frac{\int \mathcal{D}\left[ g_{ij}\right] \Psi ^{\ast }\left[
g_{ij}\right] \int_{\Sigma }d^{3}x\mathcal{H}\Psi \left[ g_{ij}\right] }{%
\int \mathcal{D}\left[ g_{ij}\right] \Psi ^{\ast }\left[ g_{ij}\right] \Psi %
\left[ g_{ij}\right] }=\frac{1}{V}\frac{\left\langle \Psi \left\vert
\int_{\Sigma }d^{3}x\hat{\Lambda}_{\Sigma }\right\vert \Psi \right\rangle }{%
\left\langle \Psi |\Psi \right\rangle }=-\frac{\Lambda }{\kappa },
\label{expect}
\end{equation}%
where we have integrated over a hypersurface $\Sigma $, divided by its
volume and functionally integrated over quantum fluctuation with the help of
some trial wave functionals. $\mathcal{H}$ is the hamiltonian constraint
without the cosmological term. Note that Eq.$\left( \ref{expect}\right) $
can be derived starting with the Wheeler-De Witt equation (WDW) \cite{DeWitt}
which represents invariance under \textit{time} reparametrization. Eq.$%
\left( \ref{expect}\right) $ represents the Sturm-Liouville problem
associated to the cosmological constant. The related boundary conditions are
dictated by the choice of the trial wavefunctionals which, in our case are
of the Gaussian type. Different types of wavefunctionals correspond to
different boundary conditions. The evaluation of the r.h.s. of Eq.$\left( %
\ref{Lind}\right) $ is performed via the absorption of the divergent part
into the re-definition of the bare classical constant $\Lambda \rightarrow
\Lambda _{0}+\Lambda ^{div}$. The dependence on the arbitrary mass scale $%
\mu $ can be removed with the help of a renormalization group-like equation.
Solving it we find that the renormalized constant $\Lambda _{0}$ should be
treated as a running one in the sense that it varies provided that the scale 
$\mu $ is changing. Finally, we obtain%
\begin{equation}
\frac{\Lambda _{0}\left( \mu _{0},r\right) }{8\pi G}=-\frac{1}{64\pi ^{2}}%
\sum_{i=1}^{2}\left[ m_{i}^{4}\left( r\right) \ln \left( \frac{%
m_{i}^{2}\left( r\right) }{4\mu _{0}^{2}}\sqrt{e}\right) \right] ,
\label{lambdaeff}
\end{equation}%
where $m_{i}^{2}\left( r\right) $ play the r\^{o}le of an effective mass with%
\begin{equation}
\left\{ 
\begin{array}{c}
m_{1}^{2}\left( r\right) =\frac{6}{r^{2}}\left( 1-\frac{b\left( r\right) }{r}%
\right) +\left[ \frac{3}{2r^{2}}b^{\prime }\left( r\right) -\frac{3}{2r^{3}}%
b\left( r\right) \right]  \\ 
\\ 
m_{2}^{2}\left( r\right) =\frac{6}{r^{2}}\left( 1-\frac{b\left( r\right) }{r}%
\right) +\left[ \frac{1}{2r^{2}}b^{\prime }\left( r\right) +\frac{3}{2r^{3}}%
b\left( r\right) \right] 
\end{array}%
\right. .  \label{potentials}
\end{equation}%
Thus, by changing the form of $b\left( r\right) $, one can explore the
spectrum of $\Lambda _{0}\left( \mu _{0},r\right) /8\pi G$. For example, we
find\cite{Remo1,Remo2}:

\begin{equation*}
\begin{tabular}{|p{3.2cm}|p{4.cm}|p{4.1cm}|}
\hline
$\text{\textrm{Schwarzschild/Naked}}$ & $b_{S}\left( r\right) =2MG$ & $%
b_{Naked-S}\left( r\right) =-2MG$ \\ \hline
&  &  \\ \hline
$\text{\textrm{dS/AdS}}$ & $b_{dS}\left( r\right) =\frac{\Lambda _{dS}}{3}%
r^{3}$ & $b_{AdS}\left( r\right) =-\frac{\Lambda _{AdS}}{3}r^{3}$ \\ \hline
&  &  \\ \hline
$\text{\textrm{SdS/S-AdS}}$ & $b_{SdS}\left( r\right) =2MG+\frac{\Lambda
_{dS}}{3}r^{3}$ & $b_{SAdS}\left( r\right) =2MG-\frac{\Lambda _{AdS}}{3}r^{3}
$ \\ \hline
\end{tabular}%
\text{.}
\end{equation*}

Note that the method represented by Eqs.$\left( \ref{p03},\ref{expect}%
\right) $ can be applied even to the electric/magnetic charge by replacing
in Eq.$\left( \ref{lambdaeff}\right) $ the induced cosmological constant
with the induced electric/magnetic charge. In doing so we obtain the quantum
gravitational contribution to such observables\cite{Remo3}%
\begin{equation}
Q_{e,0}^{2}\left( \mu _{0},r\right) =-\frac{r^{4}}{32\pi }\sum_{i=1}^{2}%
\left[ m_{i}^{4}\left( r\right) \ln \left( \frac{m_{i}^{2}\left( r\right) }{%
4\mu _{0}^{2}}\sqrt{e}\right) \right] .  \label{charge2}
\end{equation}%
Note the difference between Eq.$\left( \ref{lambdaeff}\right) $ and Eq.$%
\left( \ref{charge2}\right) .$In the latter the presence of the factor $%
r^{4} $ at the numerator allows a less singular behavior close to the
throat. A bound on $Q_{e,0}^{2}\left( \mu _{0},r\right) $ shows that%
\begin{equation}
2.\,\allowbreak 2\times 10^{-2}=\frac{9}{128\pi }\geq Q_{e,0}^{2}\left( \mu
_{0},\bar{r}\right) \geq \frac{9}{200\pi }=1.\,\allowbreak 4\times 10^{-2}.
\label{bound}
\end{equation}%
Note that the fine structure constant is $\frac{1}{137}=\allowbreak
.7\,\allowbreak 3\times 10^{-2}$. Note also that Zero Point Energy allows an
\textquotedblleft \textit{induced magnetic charge}\textquotedblright , at
least in principle. It is interesting that in all these examples except the
Schwarzschild case one could impose a running of the Newton constant $G$
instead of $\Lambda _{0}\left( \mu _{0},r\right) /8\pi G$\cite{Remo2}. An
extension of the method which considers a generic function of the curvature $%
f\left( R\right) $ has been discussed in Refs.\cite{Remo2,Remo4}. The
modifed gravity model induces a modification on $\Lambda _{0}\left( \mu
_{0},r\right) /8\pi G$ which is strongly dependent on the kind of theory
considered. How this affects the induced Electric/Magnetic charge is an open
problem.

\end{document}